\documentclass[reprint,twocolumn,prb,groupedaddress,superscriptaddress,preprintnumbers,amsmath,amsfonts,amssymb,floatfix,showpacs]{revtex4-1}

\usepackage{graphicx}
\usepackage{dcolumn}
\usepackage{bm}
\usepackage{color}
\usepackage[english]{babel}

\begin{document}

\title{First-principles prediction of oxygen octahedral rotations in perovskite-structure EuTiO$_3$}

\author{Konstantin Z. {Rushchanskii}}
\affiliation{Peter Gr\"{u}nberg Institut, Quanten-Theorie der Materialien, Forschungszentrum J\"{u}lich and JARA, 52425 J\"{u}lich, Germany}
\author{Nicola A. {Spaldin}}
\affiliation{Materials Theory, Department of Materials, ETH Zurich, Wolfgang-Pauli-Strasse 27, CH-8093 Zurich, Switzerland}
\author{Marjana {Le\v{z}ai\'{c}}}
\affiliation{Peter Gr\"{u}nberg Institut, Quanten-Theorie der Materialien, Forschungszentrum J\"{u}lich and JARA, 52425 J\"{u}lich, Germany}

\date{\today}

\begin{abstract}

We present a systematic first-principles study of the structural and vibrational properties 
of perovskite-structure EuTiO$_3$. Our calculated phonon spectrum of the high symmetry cubic
structural prototype shows strong M- and R-point instabilities, indicating a tendency to 
symmetry-lowering structural deformations composed of rotations and tilts of the oxygen 
octahedra. Subsequent explicit study of 14 different octahedral tilt-patterns showed that
the $I4/mcm$, $Imma$ and $R\bar{3}c$ structures, all with antiferrodistortive rotations of 
the octahedra, have significantly lower total energy than the prototype $Pm\bar{3}m$ structure.
We discuss the dynamical stability of these structures, and the influence of the antiferrodistortive 
structural distortions on the vibrational, optical and magnetic properties of EuTiO$_3$, in 
the context of recent unexplained experimental observations. 
\end{abstract}


\maketitle

\section{Introduction}
\label{sec:introduction}

Europium titanate, EuTiO$_3$,  was the first ternary compound of divalent Eu$^{2+}$ to be identified. \cite{McGuire:1966,mccarthy_1969}
It is a G-type antiferromagnet (AFM) with the low magnetic ordering temperature of 5.3~K, consistent with the 
highly localized $4f$ electrons on the Eu$^{2+}$ ions. \cite{McGuire:1966,Chien:1974,Chae_pss_2009} 
Until recently, its structural ground state was thought to be the ideal cubic perovskite structure, with
$Pm\bar{3}m$ symmetry. \cite{mccarthy_1969}
EuTiO$_3$ is insulating with a high dielectric constant ($\sim400$) at low temperature, indicating quantum paraelectric 
behavior and proximity to ferroelectric instability. \cite{Katsufuji:2001}
Strong interactions between the magnetic and dielectric properties have been reported \cite{Katsufuji:2001,PSSB:PSSB200402091,jiang:2121,Wu2005487,Wu2004358} including evidence for an unusual third-order coupling. \cite{PhysRevB.81.064426}
It has been suggested that EuTiO$_3$ is the prototype for studying quantum paraelectric behavior in magnetic systems. \cite{PhysRevB.81.024102} 

In addition to these experiments on bulk EuTiO$_3$, there has been a series of recent studies on EuTiO$_3$ thin films
\cite{fujita:062512,wang:5324,lee:212509,Kugimiya20072268} motivated in part by the subsequently verified 
prediction that epitaxial strain should induce ferroelectricity and multiferroicity. \cite{Fennie_ETO_2006,ETO_nature}
Likewise ''chemical strain'', {\textit i.e.} doping at the A-site with larger divalent ions, has been shown to induce 
ferroelectricity, e.g. in alloys with BaTiO$_3$. \cite{PhysRevB.69.014104}
The latter is particularly intriguing, as multiferroic (Eu,Ba)TiO$_3$ ceramics fulfil the requirements for a
solid state search for the electric dipole moment of electron. \cite{EDM_NatMat_2010} 

In this work we re-visit bulk EuTiO$_3$, with a particular focus on the possibility that the ground state
is not, in fact, the high symmetry $Pm\bar{3}m$ perovskite protoype, but a lower symmetry structure possessing
rotations or tilts of the oxygen octahedra. Until recently, all computational studies of EuTiO$_3$ 
had neglected this possibility. \cite{Fennie_ETO_2006,Ranjan_2007,ranjan:053905}
We mentioned recently in Ref.~\onlinecite{EDM_NatMat_2010} that in the course of calculating the properties of
(Eu,Ba)TiO$_3$ we noticed a calculated antiferrodistorted ground state for EuTiO$_3$.
Then in Ref.~\onlinecite{ETO_nature} the influence of such possible oxygen octahedra tiltings on the induced polarization in 
strained EuTiO$_3$ was discussed. Finally, a recent study \cite{ABH_2011} confirmed experimentally the presence of a
structural phase transition in EuTiO$_3$ using specific heat measurements, and suggested, in analogy to SrTiO$_3$, that 
the low-temperature phase should have oxygen rotations and $I4/mcm$ symmetry.

\begin{figure}[!]
\includegraphics[width=0.95\hsize]{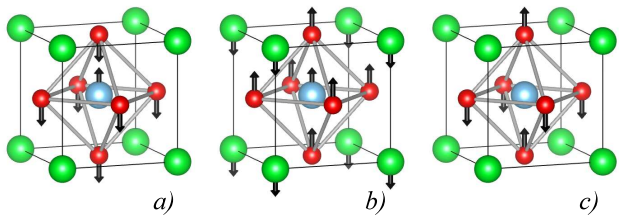}
\caption{\label{fig:perovskite_modes} (Color online) Schematic diagrams \cite{Hlinka_review} of the atomic displacements in (a) Slater mode, in which the B-site cation moves in the opposite direction from the rigid oxygen octahedral cage; (b) Last mode, in which the A-site cation moves opposite to the rigid BO$_6$ cage, and (c) Axe mode, in which the apical oxygens move opposite to the in-plane oxygens and the A- and B-site cations do not participate; a deformation of the oxygen cage results.
}
\end{figure}

Our present \textit{ab initio} study of EuTiO$_3$ was motivated by a series of experimental papers, \cite{Kamba_EPL_2007, Kamba_ETO_2009}
in which infrared (IR) reflectivity and time-domain terahertz transmission spectra of EuTiO$_3$ ceramics 
yielded intriguing results that can be better understood if the crystalline symmetry is lower then $Pm\bar{3}m$. 
The studies \cite{Kamba_EPL_2007, Kamba_ETO_2009} found three polar optical phonons between $6-600$~K, typical for the perovskite structure.
Analysis of the oscillator strengths (or equivalently, the mode-plasma frequencies) and comparison of their values with those of known perovskite prototypes such as PbTiO$_3$ (with ferroelectrically active A- and B-site cations) and BaTiO$_3$ (with active B-site cations) suggested that 
(i) The lowest-energy transverse-optical (TO) phonon TO1 corresponds predominantly to the Slater mode, 
\cite{Hlinka_review} with opposite vibration of the Ti cations and the oxygen octahedra
(Fig.~\ref{fig:perovskite_modes} left);
softening of this mode dominates the quantum paraelectric behavior,
(ii) The second lowest-energy TO2 phonon consists of vibrations of the Eu cation against the TiO$_6$ 
octahedra, commonly known as the Last mode,
(Fig.~\ref{fig:perovskite_modes} center), 
and (iii) The highest frequency TO4 (the TO3 mode is not IR active) phonon represents oxygen octahedral 
bending, usually called an Axe mode (Fig.~\ref{fig:perovskite_modes} right).

However, some features observed in the experiment are not understood: 
(i) A significant temperature dependence of the oscillator strength was found for the TO1 and TO2 modes, 
indicating changes in the vibrational mode eigenvectors with temperature. The changes are most apparent 
in the temperature range $250-400$~K (see Figure~5 in Ref.~\onlinecite{Kamba_ETO_2009}), with the
oscillator strength for the TO1 mode decreasing and for the TO2 mode increasing with increasing temperature.
(ii) The width of the peak in the imaginary dielectric function $\epsilon''$ that is associated with the 
soft TO1 mode increases when temperature is lowered. This is unusual, because at low temperature anharmonic 
effects that could give rise to the widening of the peak in $\epsilon''$ usually decrease (see Figure~3(b) 
in Ref.~\onlinecite{Kamba_ETO_2009}).
(iii) The mode at around 420~cm$^{-1}$ in the IR reflectivity spectrum is temperature dependent (see Figure~2 
in Ref.~\onlinecite{Kamba_ETO_2009}). However the Eu$_2$Ti$_2$O$_7$ pyrochlore phase, to which this peak
was tentatively assigned, does not show a structural phase transition. 

In this work we present a detailed density-functional theory based study of the structural and vibrational
properties of antiferromagnetic EuTiO$_3$. Our main finding is that the ground state is not in fact the
high symmetry $Pm\bar{3}m$ structure (in analogy with other perovskite  materials, for example KNbO$_3$\cite{PhysRevB.54.2421}), but a lower symmetry phase with tilts and rotations of the oxygen 
octahedra. The calculated properties of the lower symmetry phase are more consistent with the measured 
behavior than those calculated for the high symmetry $Pm\bar{3}m$ structure. In addition, we obtain a 
strong magneto-phonon interaction in our new ground-state structure. 
The remainder of this paper is organized as follows: In Section~\ref{sec:details} we describe the technical 
details of our calculations; Section~\ref{sec:results} contains the results of our calculations of the
structural and dynamical properties of EuTiO$_3$, and a new interpretation of the IR optical properties 
along with the influence of magnetic fields. We conclude in Section~\ref{sec:conclusion}.

\section{Calculation details}
\label{sec:details}
We carried out first-principles density-functional calculations within the spin-polarized generalized gradient approximation (GGA) \cite{PBE:1996}. For the electronic structure calculations and structural relaxations we used projector augmented-wave potentials as implemented in Vienna \textit{Ab initio} Simulation Package (VASP). \cite{VASP_Kresse:1993, VASP_Kresse:1996, Bloechl:1994, VASP_Kresse:1999} We considered the following valence-electron configuration: $5s^{2}5p^{6}4f^{7}6s^{2}$ for Eu, $3s^{2}3p^{6}3d^{2}4s^{2}$ for Ti, and $2s^{2}2p^{4}$ for oxygen.

To account for the strong electron correlation effects on the $f$-shells of Eu atoms, we used the DFT+U scheme \cite{Anisimov_et_al:1997} in Dudarev's approach \cite{Dudarev} with an on-site Coulomb parameter U=5.7~eV and Hund's exchange J$_H$=1.0~eV  after Ref.~\onlinecite{ETO_nature}.
We have checked our main conclusions (i.e. the stability of the antiferrodistorted phases relative to the cubic phase) 
for several values of these parameters (U = 4 to 8 eV) as well as without the LDA+U correction and obtained the same result.
We used a kinetic energy cutoff of 500~eV and a $6 \times 6 \times 6$ ($4 \times 4 \times 4$) $\Gamma$-centered $k$-point mesh for the unit cell (supercell) simulations. Spin-orbit interaction was not taken into account. The experimental lattice parameter of the cubic phase of EuTiO$_3$, 3.90~\AA, \cite{McGuire:1966, Chien:1974, Katsufuji:2001} was used instead of the theoretical value of 3.943~\AA. 
For structural relaxations which involved changes of the unit cell shape we performed a fixed-volume relaxation to suppress the known GGA overestimation of the lattice volume. In the structural relaxations we minimized Hellman-Feynman forces to 0.5 meV/\AA.
To investigate the structural stability, we utilized $2\times2\times2$ supercells containing 40 atoms. This allowed us to investigate all possible structural distortions which can originate from the zone-boundary instabilities in a systematic way and to minimize the error in total energy calculations when comparing energies of the different structures. 
For each symmetry investigated, the initial ideal perovskite structure was distorted according to the corresponding oxygen octahedral rotation pattern, and then the internal atomic positions and lattice parameters were relaxed until the convergence criteria were reached. 

The calculations of the dynamical properties were performed using the force-constant method. \cite{Kunc:1982, Alfe:2009}
To investigate possible dynamical instabilities we used a $2\times2\times2$ supercell of the primitive 5-atom
EuTiO$_3$ unit cell; this supercell size allows most common oxygen octahedra tilting and rotational patterns, 
as well as the likely commensurate AFM magnetic orders. 
The Hellman-Feynman forces were calculated for displacements of atoms of up to 0.04~\AA. 
The dynamical matrix for each $q$-point in the Brillouin zone was constructed by Fourier transforming the force 
constants calculated at the $\Gamma$-point and the Brillouin zone boundaries. Phonon-mode frequencies and atomic 
displacement patterns for each $q$-point were obtained as eigenvalues and eigenvectors of the dynamical matrices.
Born effective charges for symmetry-inequivalent ions in the simulation cell \cite{Ghozes:1998, Ghosez:1999} were calculated with the Berry-phase technique. \cite{King-Smith:1993}

\section{Results and discussion}
\label{sec:results}
\subsection{Properties of the high-symmetry $Pm\bar{3}m$ structure} 
In the following we show that assuming $Pm\bar{3}m$ is the ground state structure leads to discrepancies with existing experiments in the
nature of the long-wavelength phonon modes and in the mode-plasma frequencies. Because of the notorious uncertainty in DFT lattice constants we investigate wheter these discrepancies can be removed by varying the unit cell volume. We find that a partial agreement (in the mode-plasma frequencies) can be reached, but only if the lattice parameter is increased to unphysical values at which the lattice is unstable with respect to the polar displacements. Furthermore, our analysis of the full phonon spectrum of $Pm\bar{3}m$ structure indicates strong antiferrodistortive instablities.

First, we calculated the zone-center phonon frequencies and eigenvectors in cubic EuTiO$_3$ 
(see Table~\ref{tab:MPF_phonons}) and obtained results consistent with other recent calculations. 
\cite{Fennie_ETO_2006} 
As we mentioned above, in Ref.~\onlinecite{Kamba_ETO_2009} it was suggested that the TO1 mode is mainly Slater type,
based on the observation that the measured value of the mode-plasma frequency for this mode is in the same range as 
for typical ferroelectric perovskites with active B-site sublattices such as BaTiO$_3$. \cite{Hlinka_review} 
In contrast to the suggestion made in Ref.~\onlinecite{Kamba_ETO_2009}, however,
our analysis of the mode eigenvectors shows that the low-energy TO1 mode has a rather high contribution from
Eu displacements, rather than being a straightforward Slater mode.
In Table~\ref{tab:mode_decomposition} we decompose our calculated eigendisplacements of atoms in the TO1 and 
TO2 modes (calculated at the experimental lattice constant) into the Slater, Last and Axe modes discussed 
in Ref.~\onlinecite{Kamba_ETO_2009}. In addition to providing a comparison with the earlier interpretation, 
such a decomposition provides information on the relative activities of the A- and B-site cations, and 
correspondingly the role of the two cationic sublattices, in the ferroelectric transition.
Our analysis shows that both the TO1 and TO2 eigenvectors have in fact similar contributions from the 
Slater and Last modes. 
For the TO1 mode we find that the Slater- and Last-like displacements are in phase, whereas they are out of phase 
for the TO2 mode. This leads to a significant difference in the observed, as well as calculated (see below), values of the mode-plasma frequencies for both modes.

\begin{table}[!]
\caption{\label{tab:mode_decomposition}. Slater, Last, and Axe mode decomposition (in \%) of the soft polar mode TO1 and second energy polar mode TO2 for two lattice parameters of cubic EuTiO$_3$ perovskites.}
\begin{ruledtabular}
\begin{tabular}{ccccc}
Mode & $\omega$, cm$^{-1}$   & Slater & Last & Axe            \\ \hline 
\multicolumn{5}{l}{Cell parameter $a=3.90$~\AA:}   \\ 
TO1 &60     & 60 & 37 & 3                \\
TO2 &153    & 61 & 38 & 1                \\ \\ 
\multicolumn{5}{l}{Cell parameter $a=3.95$~\AA:}   \\ 
TO1 &96$i$    & 88 & 12 &                  \\
TO2 & 123   & 27 & 72 & 1                \\ 
\end{tabular}
\end{ruledtabular}
\end{table}

Next, to compare with the experimental information obtained from IR-reflectance spectra, 
\cite{Kamba_ETO_2009}  we calculated the mode-plasma frequencies  for the TO1, TO2 and TO4 
modes, using the following standard definition: \cite{Hlinka_review}

\begin{equation}
 \left(\Omega_m\right)_{\alpha\alpha}=\dfrac{1}{\sqrt{\epsilon_0V_0}}\left|\sum_{k\beta}\left(Z^*_k\right)_{\alpha\beta}m_k^{-\frac{1}{2}}\left(x_m\right)_{k\beta}\right|.
\label{MPF}
\end{equation}

\noindent Here $\left(\Omega_m\right)_{\alpha\alpha}$ is the mode-plasma frequency of mode $m$, polarized along 
direction $\alpha$, $\epsilon_0$ is the permittivity of vacuum, $V_0$ is the unit cell volume, 
$\left(Z^*_k\right)_{\alpha\beta}$ is the Born effective charge tensor of ion $k$, $m_k$ is the atomic mass 
of ion $k$, and $\left(x_m\right)_{k\beta}$ is the eigenvector of mode $m$. 
Eigenvectors are normalized to unity:
\begin{equation}
 \sum_{k\beta}\left(x_m\right)^2_{k\beta}=1.
\end{equation}
\noindent We used the Born effective charges of 2.63$|e|$ for Eu, 7.46$|e|$ for Ti, -2.18$|e|$ for O$_{\bot}$ 
and -5.74$|e|$ for O$_{||}$ which we calculated at the experimental lattice constant of 3.90 \AA\, in our 
calculation of the mode-plasma frequencies.

Interestingly, our calculated mode-plasma frequencies (Table~\ref{tab:MPF_phonons}) match well with the experimental
values in the high temperature ($300-600$~K) range, but are markedly different from the low temperature experimental 
values (see Figure~5 of Ref.~\onlinecite{Kamba_ETO_2009}). Since our DFT calculations do not include thermal
effects, particularly the strong hardening of the soft mode found with increasing temperature in EuTiO$_3$, 
we expect them instead to match the low temperature values. 
As we saw earlier, the soft mode eigenvectors
show significant volume dependence. Therefore, to check whether the discrepancy lies in our choice of
unit cell volume, we investigated the 
volume dependence of the key quantities entering Eqn.~\ref{MPF}. 
We found that the Born effective charges are largely insensitive to the volume: for values of the lattice constant 
$a$ between 3.8~\AA\ and 4.0~\AA\, the Born effective charges change from 2.68 to 2.61$|e|$ for Eu, from 7.46 
to 7.49$|e|$ for Ti, from -2.26 to -2.13$|e|$ for O$_{\bot}$, and from -5.69 to -5.83$|e|$ for O$_{||}$. 
Therefore, the changes in mode-plasma frequencies must be caused by the changes in the mode eigenvectors. 

Next, we varied the lattice parameter to see if we could obtain good agreement between the calculated and low 
temperature experimental values of the MPFs. In fact we found good agreement only by increasing $a$ to 3.95~\AA. 
From Table~\ref{tab:mode_decomposition} we can see that at $a=3.95$~\AA\ the TO1 mode is predominantly of Slater-type, while the TO2 
mode has Last character. \cite{[{Note, that one suggestion explaining the origin of the mixing of the TO1 and TO2 modes is a superexchange mechanism between 
 Eu$^{2+}$ 4f spins via the 3d states of nonmagnetic Ti$^{4+}$, see }] PhysRevB.83.214421} 
However, the Slater mode becomes imaginary, indicating a polar structural instability 
not observed experimentally in EuTiO$_3$ under normal conditions.
Moreover, increasing the lattice parameter to 3.95~\AA\ is
clearly not a physically reasonable choice -- the lattice constant at low temperature (our DFT results correspond to 0~K)
should be {\it smaller} than that at room or higher temperature because of thermal expansion of the lattice.

Therefore, we conclude that the observed discrepancy between the measured low-temperature MPFs and the calculated values 
is not due to an inappropriate choice of the lattice parameter in our calculations. Next we explore further the low-temperature 
structural properties of EuTiO$_3$ in attempt to explain the discrepancy.

\begin{table*}[!]
\caption{\label{tab:MPF_phonons} Experimental and theoretical values of the phonon wavenumbers and mode-plasma frequencies (both quantities in cm$^{-1}$) for the crystalline structures of EuTiO$_3$ with $Pm\bar{3}m$, $I4/mcm$ and $R\bar{3}c$ space groups, with ferromagnetic (FM) and G-type antiferromagnetic (AFM) magnetic ordering on the A-site cations. Mode-plasma frequency values are given in italics.
TO1, TO2 and TO4 label the optically active modes in the ideal perovsite structure; these modes are split in the tilted structures.}
\begin{ruledtabular}
\begin{tabular}{ccrrrrrrrrrrrrrrrrrrrr}
No.   & Mode & & \multicolumn{4}{c}{Exp.\footnote{Experimental data from Ref.~\onlinecite{Kamba_ETO_2009} }} & & \multicolumn{4}{c}{$Pm\bar{3}m$}   & & \multicolumn{4}{c}{$I4/mcm$}  & & \multicolumn{4}{c}{$R\bar{3}c$}\\     
   &  &&\multicolumn{2}{c}{10~K} & \multicolumn{2}{c}{400~K}   && \multicolumn{2}{c}{AFM} & \multicolumn{2}{c}{FM} && \multicolumn{2}{c}{AFM}  & \multicolumn{2}{c}{FM} &&
       \multicolumn{2}{c}{AFM} & \multicolumn{2}{c}{FM}\\ \hline  
   1& TO1    &&82 & {\it 1550} & 125 & {\it 1250} & &  67  &{\it 1289} &  60  &{\it 1327}  & & 107 &{\it 1323} & 101  &{\it  1394} & &  80 &{\it  1300} &   72 &{\it  1353} \\
   2&       &&&                &     &            & &      &           &      &            & & 128 &{\it 1607} & 118  &{\it  1622} & & 124 &{\it  1500} &  117 &{\it  1540} \\ \\
   3& TO2    &&153& {\it 400}  & 159 & {\it 860 } & &  155 &{\it 925 } & 153  &{\it  864}  & & 154 &{\it  898} & 153  &{\it   813} & & 153 &{\it   565} &  152 &{\it   452} \\
   4&       &&&                &     &            & &      &           &      &            & & 156 &{\it  314} & 156  &{\it   239} & & 160 &{\it   984} &  157 &{\it   911} \\ \\
   5&       &&&                &     &            & &      &           &      &            & & 164 &           & 218  &            & & 164 &            &  147 &            \\ \\
   6&       &&&                &     &            & &      &           &      &            & & 251 &{\it   40} & 251  &{\it    45} & & 242 &{\it   200} &  241 &{\it   188} \\ 
   7&       &&&                &     &            & &      &           &      &            & & 419 &{\it  250} & 419  &{\it   242} & & 419 &{\it   233} &  419 &{\it   241} \\ \\
   8& TO4    &&539& {\it 619}  & 540 & {\it 625 } & & 537  &{\it 824 } & 536  &{\it  820}  & & 523 &{\it  732} & 522  &{\it   729} & & 514 &{\it   737} &  513 &{\it   732} \\ 
   9&       &&&                &     &            & &      &           &      &            & & 531 &{\it  718} & 530  &{\it   715} & & 533 &{\it   733} &  532 &{\it   728} \\ \\
  10&       &&&                &     &            & &      &           &      &            & & 802 &           & 805  &            & & 803 &            &  804 &            \\
    \end{tabular}
\end{ruledtabular}
\end{table*}

\begin{table}[!]
\caption{\label{tab:MPF_phonons_Imma} Phonon wavenumbers and mode-plasma frequencies (both quantities in cm$^{-1}$) for the orthorhombic $Imma$ phase of EuTiO$_3$, with ferromagnetic (FM) and antiferromagnetic (AFM) ordering on the A-site cations. Mode-plasma frequency values are given in italics.
The meaning of the TO1, TO2 and TO4 labels is the same as in Table~\ref{tab:MPF_phonons}.}
\begin{ruledtabular}
\begin{tabular}{ccrrrrrr}
 No.   & Mode &   &\multicolumn{2}{c}{AFM}& & \multicolumn{2}{c}{FM} \\ \hline
   1   &      &   &      36 &             & & 34 &            \\
   2   &      &   &      40 &             & & 49 &            \\ \\
   3   &TO1   &   &      98 &{\it 1318   }& & 85 &{\it 1414}  \\
   4   &      &   &     104 &             & &104 &            \\
   5   &      &   &     106 &             & &105 &{\it 1479}  \\
   6   &      &   &     110 &             & &107 &            \\
   7   &      &   &     110 &{\it  1233  }& &112 &            \\
   8   &      &   &     131 &{\it  1596  }& &119 &{\it 1609}  \\ \\
   9   &TO2   &   &     154 &{\it   822  }& &152 &{\it  656}  \\
  10   &      &   &     155 &{\it   201  }& &155 &{\it  816}  \\
  11   &      &   &     159 &{\it   945  }& &156 &{\it  107}  \\ \\
  12   &      &   &     164 &             & &166 &            \\ \\
  13   &      &   &     237 &{\it   134  }& &235 &{\it  134}  \\
  14   &      &   &     249 &             & &247 &            \\
  15   &      &   &     283 &{\it   122  }& &282 &{\it   76}  \\ \\
  16   &      &   &     413 &             & &412 &            \\
  17   &      &   &     414 &             & &415 &            \\ \\
  18   &      &   &     416 &{\it   224  }& &418 &{\it  167}  \\
  19   &      &   &     418 &{\it   230  }& &422 &{\it  302}  \\ \\
  20   &      &   &     424 &             & &424 &            \\
  21   &      &   &     436 &             & &434 &            \\
  22   &      &   &     491 &             & &486 &            \\
  23   &      &   &     499 &             & &498 &            \\ \\
  24   &TO4   &   &     516 &{\it   730  }& &513 &{\it  716}  \\
  25   &      &   &     523 &{\it   736  }& &524 &{\it  743}  \\
  26   &      &   &     537 &{\it   725  }& &535 &{\it  707}  \\ \\
  27   &      &   &     801 &             & &802 &            \\
\end{tabular}
\end{ruledtabular}
\end{table}

\begin{figure}[!]
\includegraphics[width=0.95\hsize]{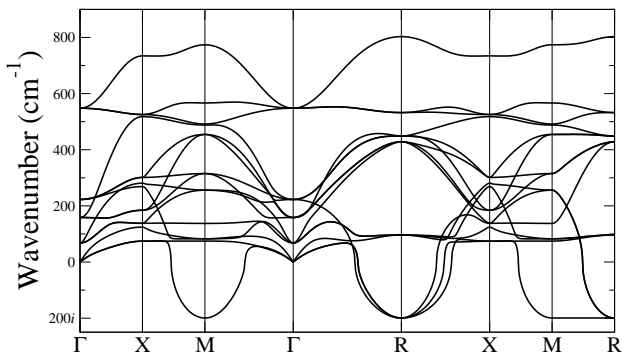}
\caption{\label{fig:phonon_Pm3m} Phonon spectrum of ideal cubic perovskite EuTiO$_3$, 
calculated at the experimental volume with lattice parameter $a=3.90$~\AA. The imaginary
wavenumbers of the low-lying modes at the Brillouin zone boundary M- and R- points indicate
structural instabilities. 
Ferromagnetic ordering of Eu 
magnetic moments was used for computational convenience. 
}
\end{figure}

Our calculated phonon spectrum of EuTiO$_3$ in the $Pm\bar{3}m$ phase (Fig.~\ref{fig:phonon_Pm3m}) indeed
reveals strong structural instabilities at the R ($\frac{1}{2},\frac{1}{2},\frac{1}{2}$) and M 
($\frac{1}{2},\frac{1}{2},0$) high symmetry points, and along the M-R symmetry line.
These instabilities are seen as a modes with imaginary wavenumbers and 
are similar to those found in calculations for $Pm\bar{3}m$ SrTiO$_3$ \cite{STO_abinitio_phonon} 
which is known experimentally to have the $I4/mcm$ ground state with alternating rotations of the oxygen octahedra
around the $c$ axis (Glazer notation \cite{Glazer:a09401} $a^0a^0c^-$). As in the case of SrTiO$_3$, our calculated 
eigenvectors indicate that these instabilities are non-polar, and arise from the tilting and rotation
of the oxygen octahedra. The eigenvector for the M-point instability shows in-phase rotations of
the oxygen octahedra around one or more pseudocubic axes, 
whereas at the R point, octahedra rotate with alternating out-of-phase sense. 

\subsection{Search for the structural ground state}

Motivated by our finding of rotational instabilities in ideal cubic perovskite EuTiO$_3$, we next
search for the structural ground state by comparing the calculated total energies of structures 
containing different combinations of rotations of the TiO$_6$ octahedra.

\begin{figure}[!]
\includegraphics[width=0.75\hsize, angle=0]{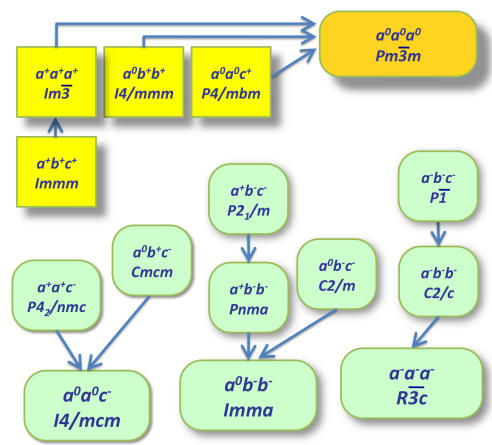}
\caption{\label{fig:diagram} (Color online) A schematic diagram, indicating the high-symmetry cubic perovskite structure and the possible 14 subgroup structures, which are the superpositions of oxygen octahedra rotations. \cite{Howard:ta0003} Superpositions are indicated by Glazer's notations together with the corresponding crystalline symmetries. 
The structures indicated by rectangles relax to ideal cubic perovskite. 
The structures indicated by the rounded rectangles represent the families of tilted structures with energies lower than that of the ideal perovskite structure. 
Arrows indicate that a structure with lower symmetry relaxed to a higher-symmetry one. For example $P2_1/m$ and its 
high-symmetry neighbour $Pnma$ structure are both unstable and relax to $Imma$. Likewise the unstable $P\bar{1}$ and 
$C2/c$ structures relax to the $R\bar{3}c$ phase. 
}

\end{figure}

As categorized by Howard and Stokes,\cite{Howard:ta0003} condensation of combinations of M- and
R-point octahedral rotations in the perovskite structure can yield any of 15 possible symmetry 
space-groups (including ideal perovskite structure), summarized in Figure~\ref{fig:diagram}. 
The following space group symmetries are allowed: $Pm\bar{3}m$, $Im\bar{3}$, $I4/mmm$, $P4/mbm$, 
$I4/mcm$, $Imma$, $R\bar{3}c$, $Immm$, $P4_2/nmc$, $Cmcm$, $Pnma$, $C2/m$, $C2/c$, $P2_1/m$ and 
$P\bar{1}$. We calculated the total energies of each of these combinations, fully relaxing the
ionic positions within the constraints of the chosen symmetry, and with the unit cell volume
set to the experimental value. The shape of the unit cell was allowed to change according to the symmetry constraints.

\begin{figure}[!]
\includegraphics[width=0.95\hsize]{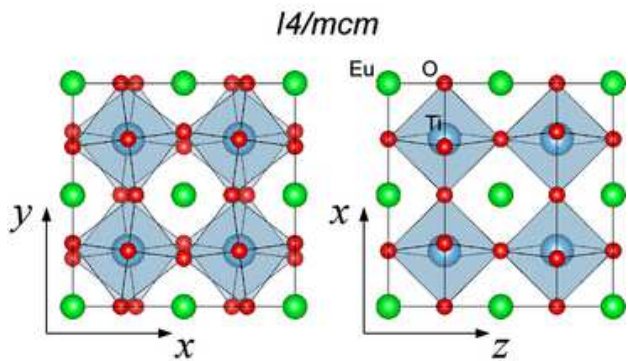}

\caption{\label{fig:Rotation_schemes} (Color online)  Schematic representation of EuTiO$_3$ in $I4/mcm$ space group structure with oxygen octahedra tilting.
Left -- view along the $z$ direction, where the out-of-phase tilting occurs, right -- view along the $y$ direction without any tilting. 
The amount of tilting corresponds to the one obtained in calculations. Drawings were produced by VESTA visualisation software.\cite{VESTA}  
}
\end{figure}

\begin{figure}[!]
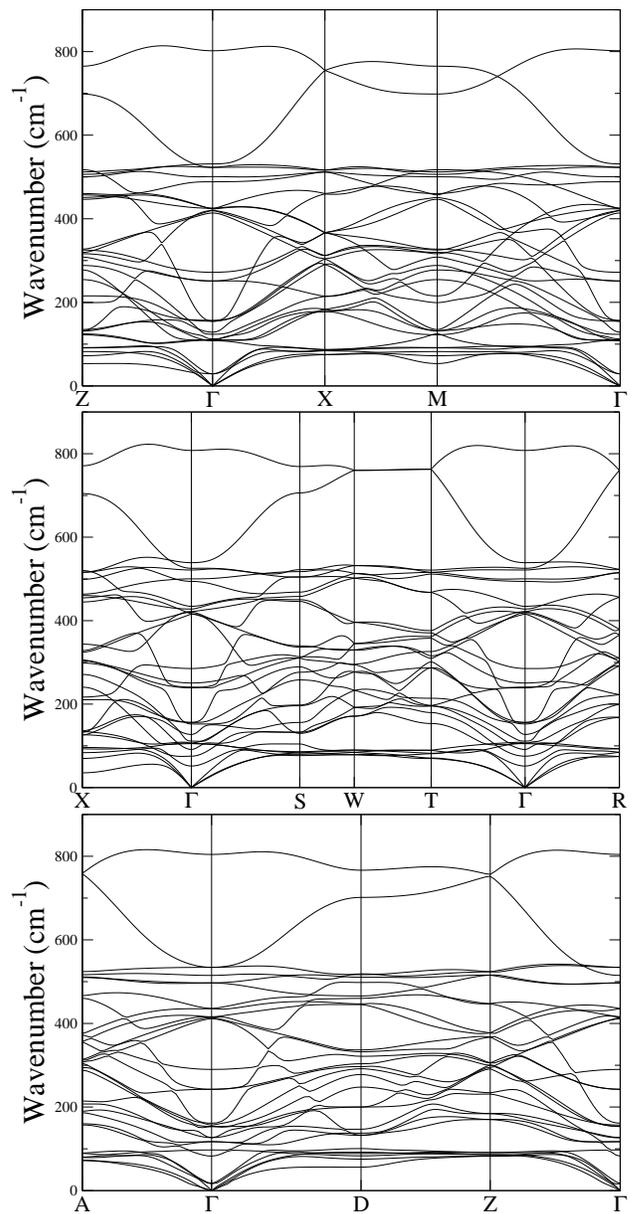

\includegraphics[width=0.95\hsize]{I4mcm_full_phonon_spectrum.eps}
\includegraphics[width=0.95\hsize]{Imma_full_phonon_spectrum.eps}
\includegraphics[width=0.95\hsize]{R3c_full_phonon_spectrum.eps}
\caption{\label{fig:phonon} Full phonon spectra of EuTiO$_3$ in $I4/mcm$ (upper panel), $Imma$ 
(middle panel) and $R\bar{3}c$ (lower panel) structures. G-type AFM ordering of Eu magnetic moments
was used in each case. All three structures are dynamically stable.
In units of the reciprocal lattice vectors, the high symmetry points are as follows. $I4/mcm$: 
X = (0,$\frac{1}{2},0$), M = ($\frac{1}{2},\frac{1}{2},0$), Z = ($\frac{1}{2},\frac{1}{2},\frac{1}{2}$); 
$Imma$: X = ($\frac{1}{2},\frac{1}{2},-\frac{1}{2}$), S = ($\frac{1}{2},0,0$), R = ($0,\frac{1}{2},0$), 
T = ($0,0,\frac{1}{2}$), W = ($\frac{1}{4},\frac{1}{4},\frac{1}{4}$); 
$R\bar{3}c$: A = ($\frac{1}{2},0,0$), D = ($\frac{1}{2},\frac{1}{2},0$), Z = 
($\frac{1}{2},\frac{1}{2},\frac{1}{2}$).
}
\end{figure}

We found that many of the low symmetry tilt patterns relaxed to higher symmetry structures, as
indicated by the arrows in Figure~\ref{fig:diagram}.  
Three structures remained stable -- those with $R\bar{3}c$, $Imma$ and $I4/mcm$ space groups. All three are
stabilized by around 25meV per formula unit compared with the prototype $Pm\bar{3}m$ structure.
The $I4/mcm$ is lowest in energy ($\sim$27 meV per formula unit lower than $Pm\bar{3}m$), $Imma$ is the second most stable structure ($\sim$26 meV per formula unit lower), then $R\bar{3}c$ at $\sim$25 meV per formula unit lower. 
G-type antiferromagnetic configuration was found to be of lower energy than the ferromagnetic one in all tilted phases.
Importantly, the energy differences between the $I4/mcm$, $Imma$ and $R\bar{3}c$ structures are very small, within 
$\sim$2~meV per formula unit, which is the range of numerical errors. Therefore, we can not predict
which rotation scenario is the ground state of EuTiO$_3$. All three structures have no imaginary phonons
through the entire Brillouin zone (see Figure~\ref{fig:phonon})
and so are dynamically stable. 
Note that the $I4/mcm$ structure is obtained by condensing out-of-phase rotations of oxygen octahedra about
only one axis (see Figure~\ref{fig:Rotation_schemes}), $Imma$ results from the same out-of-phase rotations about two axes, and $R\bar{3}c$ 
from out-of-phase rotations about all three axes. Our calculated structural parameters for all phases 
are collected in Table~\ref{tab:structure}.  

\begin{table}[!]
\caption{\label{tab:structure} Theoretical structural parameters of EuTiO$_3$ for $a^0a^0c^-$ ($I4/mcm$), $a^0b^-b^-$ ($Imma$) and $a^-a^-a^-$ ($R\bar{3}c$) oxygen octahedral rotation patterns.}
\begin{ruledtabular}
\begin{tabular}{ccddd}
 \multicolumn{5}{c}{$I4/mcm$}         \\  
       & Wyckoff      &  \multicolumn{1}{c}{\mbox{$x$}} & \multicolumn{1}{c}{\mbox{$y$}} &\multicolumn{1}{c}{\mbox{$z$}}    \\  \hline
Eu     &    2($b$)    &  0           &  0.5      & 0.25       \\
Ti     &    2($c$)    &  0           &  0        & 0          \\
O(1)   &    2($a$)    &  0           &  0        & 0.25       \\
O(2)   &    4($h$)    &  0.21578     &  0.71578  & 0          \\  \\
\multicolumn{2}{c}{Cell parameters} \\
 &  \mbox{$a$} (\AA)  & 5.494 \\
 &  \mbox{$c$} (\AA)  & 7.861 \\     
\\
\multicolumn{5}{c}{$Imma$}         \\  
       & Wyckoff      &  \multicolumn{1}{c}{\mbox{$x$}} & \multicolumn{1}{c}{\mbox{$y$}} &\multicolumn{1}{c}{\mbox{$z$}}    \\  \hline
Eu     &    4($e$)    &  0           &  0.25     & 0.00221    \\
Ti     &    4($b$)    &  0           &  0        & 0.5        \\
O(1)   &    4($e$)    &  0           &  0.25     & 0.54663    \\
O(2)   &    8($g$)    &  0.25        &  0.02415  & 0.25       \\  \\
\multicolumn{2}{c}{Cell parameters} \\
 &  \mbox{$a$} (\AA)  & 5.509\\
 &  \mbox{$b$} (\AA)  & 7.774 \\
 &  \mbox{$c$} (\AA)  & 5.541 \\
\\
\multicolumn{5}{c}{$R\bar{3}c$} \\     
       & Wyckoff      &  \multicolumn{1}{c}{\mbox{$x$}} & \multicolumn{1}{c}{\mbox{$y$}} &\multicolumn{1}{c}{\mbox{$z$}}    \\  \hline
Eu     &    2($a$)    &  0           &  0        & 0.25       \\
Ti     &    2($b$)    &  0           &  0        & 0          \\
O      &    6($e$)    &  0.53813     &  0        & 0.25       \\  \\
\multicolumn{2}{c}{Cell parameters} \\
 &  \mbox{$a$} (\AA)  & 5.528\\
 &  \mbox{$c$} (\AA)  & 13.448 \\
 
\end{tabular}
\end{ruledtabular}
\end{table}

To check whether our conclusions on the phase stability are affected by volume, we next investigated the 
athermal equation of state for the three candidate ground-state phases, as well as for the undistorted one. 
We found that the $E(V)$ dependence is adequately described by means of the first-order Murnaghan equation of state: \cite{Murnaghan}

\begin{equation}
E=E_0+\dfrac{B_0V}{B_0'}\left(\frac{(V_0/V)^{B_0'}}{B_0'-1}+1\right)-\dfrac{B_0V_0}{B_0'-1},
\end{equation}

\noindent where $B_0$ is the bulk modulus, $B_0'$ its pressure derivative, $V_0$ the equilibrium volume, and $E_0$ is the energy minimum corresponding to the equilibrium volume. Our calculated parameters for this Eqn., obtained by fitting of \textit{ab initio} total energies, are listed in Table~\ref{tab:murn}. The equilibrium  values for $V_0$, $E_0$ and bulk modulus $B_0$ for tetragonal, orthorhombic and rhombohedral phases are almost the same, however the pressure derivative $B_0'$ for $R\bar{3}c$ phase is about 1.5 times lower then for the $I4/mcm$ phase. This indicates that the internal relaxations in the rhombohedral phase are less sensitive to the volume changes than in the tetragonal phase. Also, the $R\bar{3}c$ structure is more rigid than the cubic and the tetragonal ones. 
Measurements of the bulk moduli and their pressure derivatives at different temperatures would be helpful
in distinguishing between these lower-symmetry phases.

\begin{table}[h]
\caption{\label{tab:murn} Calculated Murnaghan equation of state parameters for the $I4/mcm$, $Imma$ 
and $R\bar{3}c$ structures. $V_0$ is the theoretical volume and $a_0=(V_0)^{1/3}$ is the theoretical 
pseudo-cubic lattice constant.}
\begin{ruledtabular}
\begin{tabular}{ldddd}
                & \multicolumn{1}{c}{\mbox{$Pm\bar{3}m$}} & \multicolumn{1}{c}{\mbox{$I4/mcm$}}  & \multicolumn{1}{c}{\mbox{$Imma$}}  & \multicolumn{1}{c}{\mbox{$R\bar{3}c$}}  \\  \hline
$V_0$ (\AA$^3$) &   61.29      &  61.04    &  61.11  &  61.08       \\
$a_0$ (\AA)     &    3.943     &   3.937   &   3.939 &   3.938      \\
$B_0$ (GPa)     &  162.5       & 171.1     & 169.4   & 179.1        \\
$B_0'$          &   10.82      &   7.57    &   4.98  &   4.55       \\
\end{tabular}
\end{ruledtabular}
\end{table}

\subsection{IR optical properties in structures with tilting}

Finally, we attempt to distinguish between the three candidate ground state structures by comparing their
calculated phonon and mode-plasma frequencies with the measured optical properties. We find that the
calculated frequencies of the $R\bar{3}c$ structure show best agreement with experiment, suggesting it 
as a likely experimental structure. 

\begin{figure}[!]
\includegraphics[width=0.75\hsize]{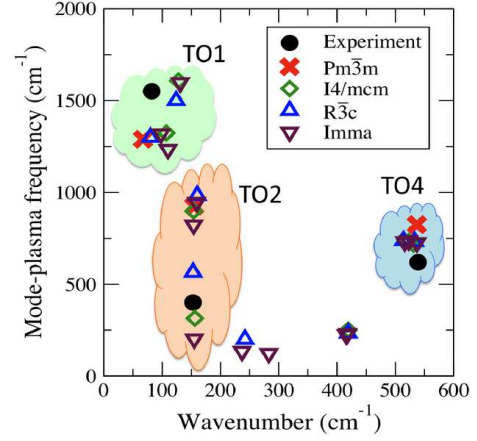}
\caption{\label{fig:MPF_compare} (Color online) Calculated and measured \cite{Kamba_ETO_2009} mode-plasma 
frequencies, plotted as a function of the mode wavenumbers. Our calculated values for four different structural
symmetries are presented. Values originating from the TO1, TO2 and TO4 modes of the ideal perovskite are 
grouped together; the ungrouped symbols correspond to additional modes that appear due to the symmetry lowering.
}
\end{figure}

Our results are summarized in Tables~\ref{tab:MPF_phonons} and \ref{tab:MPF_phonons_Imma}  as well as in 
Figure~\ref{fig:MPF_compare}. In all cases we used the Born effective charges calculated for the 
$Pm\bar{3}m$ reference structure in our evaluation of the mode plasma frequencis, and our reported values
do not include the LO-TO splitting. Calculations were performed for the G-type AFM ordered phase; the
coupling between phonon properties and magnetism is discussed at the end of this section.

First we look at the frequency of the soft TO1 mode (see Table~\ref{tab:MPF_phonons}). Our calculated 
values are 107~cm$^{-1}$ and 128~cm$^{-1}$ for $I4/mcm$ (the symmetry lowering splits the TO1 mode into one doubly
degenerate and one singly degenerate $c$ polarized mode), 80~cm$^{-1}$ and 124~cm$^{-1}$ 
($R\bar{3}c$) and 98 to 131~cm$^{-1}$ ($Imma$). 
The experimental value is 82~cm$^{-1}$, suggesting that the $R\bar{3}c$ structure is most likely.
We note, however, that the soft-mode 
frequency is very sensitive to the volume of the structure and small changes in choice of volume 
(GGA versus experimental for example) can strongly change the predicted values. 
Regarding the mode plasma frequencies of this mode, the experimental value is 1550~cm$^{-1}$, and
the average calculated values for the three structures are 1417~cm$^{-1}$ ($I4/mcm$), 
1433~cm$^{-1}$ ($R\bar{3}c$) and 1382~cm$^{-1}$ ($Imma$) 
all showing comparable agreement with the experimental value.
Note that the relatively large mode splittings that we obtain for TO1 modes in the lower symmetry
structures could explain the experimentally observed large peak width in the imaginary component 
of the complex dielectric function (see Figure~3b in Ref.~\onlinecite{Kamba_ETO_2009}) when two modes with lifted degeneracy are fitted as one mode. 

For completeness, we also list in Tables~\ref{tab:MPF_phonons} and \ref{tab:MPF_phonons_Imma} our calculated frequencies and mode plasma 
frequencies of the TO2- and TO4-like modes. There are no striking differences that provide persuasive 
evidence for one structure over another.
In the $I4/mcm$ structure, the splitting of the TO2 mode, which is caused by the symmetry lowering, is 
small (see modes~3 and~4 in Table~\ref{tab:MPF_phonons}). However the mode-plasma frequencies of the 
symmetry-split TO2 modes are significantly different: Mode~3 consists mainly of out-of-phase shifts of 
Eu and Ti atoms, without oxygen contribution, whereas the $c$-axis-polarized mode 4 has non-zero oxygen
displacements which reduce the relative shifts of the Ti atoms, resulting in a lower dipole for this mode. 
The average value of the mode-plasma frequencies for these three modes is $703$~cm$^{-1}$, which is 
significantly lower than the corresponding value in  the $Pm\bar{3}m$ structure (925~cm$^{-1}$) but still 
much higher than experimental one ($400$~cm$^{-1}$). The average value of the mode-plasma frequency for 
the TO4 mode (modes 8 and 9) is 727~cm$^{-1}$, compared with 820~cm$^{-1}$
for the reference $Pm\bar{3}m$ structure.
In the $R\bar{3}c$ phase the TO2 mode also splits by a small amount (7~cm$^{-1}$) with average 
mode-plasma-frequency of 704~cm$^{-1}$, and the TO4 mode averaged mode-plasma frequency is 734~cm$^{-1}$;
both values are similar to those of the $I4/mcm$ structure. 
Like in the $I4/mcm$ structure, mode~1 is built by Eu displacements against oxygen cage, with small in-phase contribution of Ti sublattice. 
The eigenvectors are also similar to those of the $Pm\bar{3}m$ and $I4/mcm$ structures.
In the $Imma$ structure (see Table~\ref{tab:MPF_phonons_Imma}), the wavenumbers of the TO2-group are in the range 154-159~cm$^{-1}$, 
and those of the TO4-group are in the range 516-537~cm$^{-1}$.  As in the $R\bar{3}c$ structure, 
modes in the range 237-283~cm$^{-1}$ are activated and should be observed as a doublet with a
splitting of $~50$~cm$^{-1}$.

One interesting observation is the presence of an IR-active mode with frequency $\sim$419~cm$^{-1}$ 
in all three structures (mode~7 in Table~\ref{tab:MPF_phonons} and 18-19 in Table~\ref{tab:MPF_phonons_Imma}).
This two-fold degenerate mode (it is a doublet in the low-symmetry $Imma$ structure) has a relatively 
low mode-plasma frequency of 250~cm$^{-1}$ (and, correspondingly, a low oscillator strength). The 
eigenvector of this mode shows that this vibration originates from the out-of-phase displacements of 
Ti atoms in the neighboring sublattices. However, the atomic displacements are not collinear, which 
leads to an uncompensated dipole in the plane, orthogonal to the $c$-axis. A mode with a similar 
frequency ($\sim$430~cm$^{-1}$) is observed at low temperatures in the experimental IR reflectivity 
data (see Figure~2 in Ref.~\onlinecite{Kamba_ETO_2009}). The fact that experimentally this mode is 
observable only at low temperatures gives an additional indication that it could be a fingerprint 
of the antiferrodistortive  phase transition.

To make the comparison with the available experimental data easier, we show in Fig.~\ref{fig:MPF_compare} 
the calculated and experimental mode-plasma frequencies as a function of wavenumber. Clearly the calculated
values for the $Pm\bar{3}m$ structure are strikingly different from the experimental values providing
further evidence that this is unlikely to be the ground state structure. It remains difficult to
distinguish between the $I4/mcm$,  $R\bar{3}c$ and $Imma$ structures which all show similarly good agreement
with current experimental data. Our predicted differences in the degeneracies and ordering of the 
components of the TO1 and TO2 modes, could in principle be distinguished experimentally using polarized 
optical spectroscopy on single crystals.

\subsection{Magneto-structural coupling}
Strong magneto-structural coupling and multiferroic effects were predicted previously for
$Pm\bar{3}m$ EuTiO$_3$ \cite{Fennie_ETO_2006}; here we investigate whether such behavior persists
in our newly predicted tilted ground-state structures. Indeed, we find that constraining the
spins to be aligned ferromagnetically decreases the soft-mode wavenumber by $\sim$7~cm$^{-1}$
for all three tilted structures; this is the same as previously calculated for the $Pm\bar{3}m$
structure\cite{Fennie_ETO_2006} (7~cm$^{-1}$), and close to 
the value estimated from the low-temperature IR reflectivity spectra at different magnetic fields 
(3~cm$^{-1}$). \cite{Kamba_EPL_2007}  
We obtain a decrease of about 2~cm$^{-1}$ for all modes belonging to the TO2 group. 
The frequency of the TO4 Axe mode is almost insensitive to the magnetic ordering.
The modes corresponding to tilting of the oxygen octahedra, on the other hand, show strong
magneto-phonon coupling, since they cause changes in the Eu-O-Eu angles which in turn strongly 
modify the superexchange pathways between the magnetic ions. 
These modes are listed in Table~\ref{tab:MPF_phonons} as mode~5. We find that ferromagnetic 
constraint on the Eu spins hardens this mode in the $I4/mcm$ structure from 164~cm$^{-1}$ 
to 218~cm$^{-1}$, whereas in the $R\bar{3}c$ structure it is softened by 17~cm$^{-1}$. Since 
the tilt mode is not polar, however, the dielectric function is not affected. 

Finally we mention that Shvartsman \textit{et al} \cite{PhysRevB.81.064426} recently measured an
unusual off-diagonal magnetoelectric coupling in EuTiO$_3$, which they rationalized using our
predictions of tilted ground states. In their experiments, the applied electric field
breaks inversion symmetry, allowing a Dzyaloshinskii-Moriya interaction and subsequent magnetoelectric
response analogous to that in perovskite FeTiO$_3$, which also shows magnetic A-site ions and octahedral 
tiltings. \cite{PhysRevLett.100.167203,PhysRevLett.103.047601} 

\section{Conclusions}
\label{sec:conclusion}
In summary, our density functional calculations for perovskite EuTiO$_3$ predict that its ground
state consists of tilts and rotations of the oxygen octahedra, rather than the simple cubic perovskite
structure as previously believed. By comparing the total energies of all symmetry allowed tilting
patterns, we identified three candidate ground states: $a^0a^0c^-$ ($I4/mcm$), $a^0b^-b^-$ ($Imma$) 
and $a^-a^-a^-$  ($R\bar{3}c$). We compared the calculated phonon properties with available 
experimental data for the three candidate structures to identify the most likely ground state. 
Our calculated energy differences are two small to allow us to predict which of these
phases is the ground state; indeed a phase coexistence might be possible, and additional
experiments are required for further progress.
In particular, a search for the infra-red-active mode in the range 220-290~cm$^{-1}$ is a signature 
of $R\bar{3}c$ and $Imma$ structures, but is not active in $I4/mcm$ phase. Also the magnetic-field
dependence of the mode at $\sim$164~cm$^{-1}$ would distinguish between the $I4/mcm$ and $R\bar{3}c$ 
structures.

\begin{acknowledgments}
This work was supported by the Young Investigators Group Programme of the Helmholtz Association, Germany, contract VH-NG-409 and by the ETH Z\"{u}rich. We thank S.~Kamba, R.~Hermann, J.~Hlinka, W.~Kleemann, C.~Fennie, S.~Bl\"ugel and I.~Slipukhina for discussions and gratefuly acknowledge the support of J\"{u}lich Supercomputing Centre.
\end{acknowledgments}


%

\end{document}